\documentclass[preprint,prc,aps,showpacs,showkeys,groupedaddress,floatfix]{revtex4}
\usepackage{dcolumn}
\usepackage{bm}
\usepackage{graphics}
\usepackage{graphicx}
\usepackage{epsfig}
\usepackage{amssymb}
\usepackage{amsmath}
\begin{document}
\title{Orthogonal polynomial solutions to the non-central modified Kratzer potential}
\author{F. Yasuk, I. Boztosun and A. Durmus}
\affiliation{Faculty of Arts and Sciences, Department of Physics,
Erciyes University, Kayseri, Turkey}
\begin{abstract}
We investigate the analytical solution of a new exactly solvable
non-central potential of $V(r,\theta) = D\left({\frac{r - a}{r}}
\right)^2+{\frac{\beta }{r^2\sin^2 \theta }}+{\frac{\gamma \cos
\theta }{r^2\sin^2 \theta }}$ type, which may be called as the
modified non-central Kratzer potential. The energy eigenvalues as
well as the corresponding eigenfunctions are calculated for various
values of $n$ and $m$ quantum numbers within the framework of the
Nikiforov-Uvarov and Asymtotic Iteration Methods for the $CO$
diatomic molecule as an application of this potential. In this
paper, we first present the effect of the non-central term on the
bound-state energy eigenvalues: this effect is determined explicitly
for different $n$ and $m$ quantum numbers with $\beta=\gamma$=0.0,
0.1, 1.0 and 5.0 values and the results are compared with the
findings of the modified Kratzer potential for different $n$ and $l$
quantum numbers. Then, we show that the angle-dependent non-central
part behaves like a centrifugal barrier and it reduces the depth of
the attractive potential pocket, which effects the bound-state
energy eigenvalues.
\end{abstract}
 \maketitle
\section{Introduction}
The analytical solution of the radial Schr\"{o}dinger equation is of
high importance in non-relativistic quantum mechanics since the wave
function contains all the necessary information to describe a
quantum system fully. There are only a few potentials for which the
radial Schr\"{o}dinger equation can be solved explicitly for all $n$
and all $l$. So far, many methods such as super-symmetry (SUSY)
\cite{susy,morales,gonul}, Nikiforov-Uvarov
\cite{Nikiforov,egrifes,sever,cuneyt,yasuk}, Asymptotic Iteration
Method \cite{hakan1,hakan2,karakoc,bayrak,aygun} and the Pekeris
approximation \cite{Pekeris,fluge} have been developed and applied
to solve the radial Schr\"{o}dinger equation exactly or
quasi-exactly for $l\neq 0$ within a given potential.

The Kratzer or modified Kratzer type potentials \cite{kratzer} we
consider in this paper have played an important role in the
history of the molecular and quantum chemistry and they have been
so far extensively used to describe the molecular structure and
interactions \cite{mol}. Although such central potentials have
been extensively used to describe the properties of the system
qualitatively, it is known that the dynamical properties of the
system should also be taken into account for a better description
of the system. That means, the potential should have not only the
radial but also the angle dependent parts. These dynamical
properties of the system can be taken into account by using a
non-central potential and therefore, in this paper, we aim to
present the effect of the non-central term to calculate the
non-zero angular momentum solutions of the Schr\"{o}dinger
equation. For this, we take the non-central modified Kratzer
\cite{fues} potential ($V(r,\theta ) = D\left( {\frac{r - a}{r}}
\right)^2 + {\frac{\beta }{r^2\sin^2 \theta }}  + {\frac{\gamma
\cos \theta }{r^2\sin^2 \theta }}$) and as an application, we
calculate the energy eigenvalues of the $CO$ diatomic molecule for
various $n$ and $m$ quantum numbers.

The article is organized as follows: In the following two sections,
the basic equations of the Nikiforov-Uvarov method (NU) as well as
the Asymptotic Iteration Method (AIM) used so far to solve the
resulting differential equation are given briefly. The solutions of
the Schr\"{o}dinger equation with the non-central modified Kratzer
potential are outlined in the fourth section. Then, the solutions of
the Schr\"{o}dinger equation with the non-central modified Kratzer
potential are obtained by using both methods. The energy eigenvalues
and the corresponding eigenfunctions are calculated for various
values of $n$ and $m$ quantum numbers for the $CO$ diatomic molecule
and the effect of the non-central term is determined explicitly by
comparing the results with the findings of the modified Kratzer
potential with different $n$ and $l$ quantum numbers. The summary
and conclusion are provided in section~\ref{conc}.

\section{Nikiforov-Uvarov Method}\label{sec2}
The NU method is based on the solutions of general second order
linear equations with special functions. It has been extensively
used to solve the non-relativistic Schr\"{o}dinger equation or
similar time-independent second-order differential equations and
there is an extensive literature to look at
\cite{Nikiforov,sever,egrifes,cuneyt,yasuk}. However, in order to
keep the completeness of the paper, we briefly outline NU method
here. The idea in the NU method is to convert the Schr\"{o}dinger
equation or similar differential equations into the following
form:
\begin{equation}
\label{eq11} \psi (s{)}'' + \frac{\tilde {\tau }(s)}{\sigma
(s)}{\psi }'(s) + \frac{\tilde {\sigma }(s)}{\sigma ^2(s)}\psi (s)
= 0
\end{equation}
where $\sigma (s)$ and $\tilde {\sigma }(s)$ are polynomials, at
most second-degree, and $\tilde {\tau }(s)$ is a first-degree
polynomial \cite{Nikiforov,egrifes,sever,cuneyt,yasuk}. Hence,
from Eq. (\ref{eq11}), the Schr\"{o}dinger equation or the
Schr\"{o}dinger-like equations can be solved analytically by this
method. In order to find a particular solution of Eq.
(\ref{eq11}), the following transformation is used:
\begin{equation}
\label{eq12} \psi (s) = \phi (s)y(s)
\end{equation}
it reduces Eq. (\ref{eq11}) to an equation of hypergeometric type,
\begin{equation}
\label{eq13} \sigma (s){y}'' + \tau (s){y}' + \lambdabar y = 0
\end{equation}
and $\phi (s)$ is defined as a logarithmic derivative in the
following form and its solutions can be obtained from
\begin{equation}
\label{eq14} {{\phi }'(s)} \mathord{\left/ {\vphantom {{{\phi
}'(s)} \phi }} \right. \kern-\nulldelimiterspace} \phi (s) = {\pi
(s)} \mathord{\left/ {\vphantom {{\pi (s)} {\sigma (s)}}} \right.
\kern-\nulldelimiterspace} {\sigma (s)}
\end{equation}
The other part $y(s)$ is the hypergeometric type function whose
polynomial solutions are given by the Rodrigues relation
\begin{equation}
\label{eq15} y_n (s) = \frac{B_n }{\rho (s)}\frac{d^n}{ds^n}\left[
{\sigma ^n(s)\rho (s)} \right]
\end{equation}
where $B_n $ is the normalization constant and the weight function
$\rho (s)$ must satisfy the condition
\begin{equation}
\label{eq16} \left( {\sigma {\kern 1pt} \rho } \right)^\prime =
\tau {\kern 1pt} \rho
\end{equation}
The function $\pi $ and the parameter $\lambdabar $ required for
this method are defined as follows
\begin{equation}
\label{eq17} \pi (s) = \frac{{\sigma }' - \tilde {\tau }}{2}\pm
\sqrt {\left( {\frac{{\sigma }' - \tilde {\tau }}{2}} \right)^2 -
\tilde {\sigma } + k\sigma } \quad ,
\end{equation}
\begin{equation}
\label{eq18} {\lambdabar} = k + {\pi }'
\end{equation}
On the other hand, in order to find the value of $k$, the
expression under the square root must be the square of a
polynomial. Thus, a new eigenvalue equation for the
Schr\"{o}dinger equation becomes
\begin{equation}
\label{eq19} \lambdabar = \lambdabar_n = - n{\tau }' - \frac{n(n -
1)}{2}{\sigma }''
\end{equation}
where
\begin{equation}
\label{eq20} \tau (s) = \tilde {\tau }(s) + 2\pi (s)
\end{equation}
and its derivative is negative. By comparison of Eqs. (\ref{eq18})
and (\ref{eq19}), we obtain the energy eigenvalues.

\section{The Asymptotic Iteration Method (AIM)}\label{aim}
In this section we briefly outline the asymptotic iteration method,
the details can be found in references
\cite{hakan1,hakan2,karakoc,bayrak,aygun}. The asymptotic iteration
method was proposed to solve second-order differential equations of
the form
\begin{equation}\label{diff}
  y''=\lambda_{0}(x)y'+s_{0}(x)y
\end{equation}
where s$_{0}$(x), $\lambda_{0}$(x) are functions in
C$_{\infty}$(a,b). The variables, s$_{0}$(x) and $\lambda_{0}$(x),
are sufficiently differentiable.

\begin{equation}
{y}''' = \lambda _1 (x){y}' + s_1 (x)y
\end{equation}
The second derivative of Eq.(\ref{diff}) is obtained as

\begin{equation}
{y}'''' = \lambda _2 (x){y}' + s_2 (x)y
\end{equation}
where

\begin{equation}
\lambda _1(x) = \lambda _{0}^{'}(x) + s_0(x) + \lambda _{0}^{2}(x)
\hspace{1cm} s_1(x)=s_{0}^{'}(x) + s_0(x)\lambda _0(x)  \nonumber \\
\end{equation}

\begin{equation}
\lambda _2(x) = \lambda _{1}^{'}(x) + s_1(x) + \lambda _0(x)
\lambda _1 \hspace{1cm} s_2(x) = s_{1}^{'}(x) + s_0(x) \lambda
_1(x)
\end{equation}
Therefore, for $(n+1)^{th}$ and $(n+2)^{th}$ derivatives,
$n$=1,2,...., one can get,
\begin{equation}
y^{(n + 1)} = \lambda _{n - 1} (x){y}' + s_{n - 1} (x)y
\end{equation}
and
\begin{equation}
y^{(n + 2)} = \lambda _n (x){y}' + s_n (x)y
\end{equation}
respectively, where
\begin{equation}\label{iter}
\lambda _n(x) = \lambda _{n - 1}^{'}(x) + s_{n - 1}(x) + \lambda
_0(x) \lambda _{n - 1}(x) \hspace{1cm}s_n(x) = s_{n - 1}^{'}(x) +
s_0(x) \lambda _{n - 1}(x)
\end{equation}
The ratio of the $(n+2)^{th}$ and $(n+1)^{th}$ derivatives can be
expressed as

\begin{equation}\label{as1}
\frac{d}{dx}\ln (y^{(n + 1)}) = \frac{y^{(n + 2)}}{y^{(n + 1)}} =
\frac{\lambda _n ({y}' + \textstyle{{s_n } \over {\lambda _n
}}y)}{\lambda _{n - 1} ({y}' + \textstyle{{s_{n - 1} } \over
{\lambda _{n - 1} }}y)}
\end{equation}
For sufficiently large $n$, we can now introduce the asymptotic
aspect of the method; that is,
\begin{equation}\label{itercond}
\frac{s_n }{\lambda _n } = \frac{s_{n - 1} }{\lambda _{n - 1} } =
\alpha
\end{equation}
Thus Eq.(\ref{as1}) can be reduced to
\begin{equation}
\frac{d}{dx}\ln (y^{(n + 1)}) = \frac{\lambda _n }{\lambda _{n -
1} }
\end{equation}
which yields the general solution of Eq.(\ref{diff}) \cite{hakan1}

\begin{equation}\label{generalsolution}
  y(x)=exp \left( - \int^{x} \alpha dt\right ) \left [C_{2}+C_{1}
  \int^{x}exp  \left( \int^{t} \lambda_{0}(\tau)+2\alpha(\tau) d\tau \right ) dt \right
  ]
\end{equation}
For a given potential such as non-central modified Kratzer,
Schr\"{o}dinger equations are converted to the form of
Eq.(\ref{diff}). Then, s$_{0}(x)$ and $\lambda_{0}(x)$ are
determined and s$_{n}(x)$ and $\lambda_{n}(x)$ parameters are
calculated. The energy eigenvalues are obtained by the termination
condition given by Eq.(\ref{itercond}).

In this study, we investigate the exact solutions of the
Schr\"{o}dinger equations which the relevant second order
homogenous linear differential equation takes the following
general form \cite{hakan2},
\begin{equation}\label{diff1}
 {y}'' = 2\left( {\frac{ax^{N + 1}}{1 - bx^{N + 2}} -
\frac{\left( {m + 1} \right)}{x}} \right){y}' - \frac{wx^N}{1 -
bx^{N + 2}}y
\end{equation}
If this equation is compared to Eq.(\ref{diff}), it entails the
following expressions
\begin{equation}\label{snln}
\lambda _0(x) = 2\left( {\frac{ax^{N + 1}}{1 - bx^{N + 2}} -
\frac{\left( {m + 1} \right)}{x}} \right)   \hspace{1cm}  s_0 (x)
= - \frac{wx^N}{1 - bx^{N + 2}}
\end{equation}
$a$, $b$ and $m$ are real numbers and $w_n^m(N)$ can be determined
from condition Eq.(\ref{itercond}) as follows
\begin{eqnarray}
w_n^m (-1) & = & n\left( {2a + 2bm + (n + 1)b} \right)\\
w_n^m (0) & = & 2n\left( {2a + 2bm + (2n + 1)b} \right) \\
w_n^m (1) & = & 3n\left( {2a + 2bm + (3n + 1)b} \right) \\
w_n^m (2) & = & 4n\left( {2a + 2bm + (4n + 1)b} \right) \\
w_n^m (3) & = & 5n\left( {2a + 2bm + (5n + 1)b} \right) \\
\ldots \emph{etc} \nonumber
\end{eqnarray}
Hence, these formulae are easily generalized as;
\begin{equation}
w_n^m (N) = b\left( {N + 2} \right)^2n\left( {n + \frac{\left( {2m
+ 1} \right)b + 2a}{\left( {N + 2} \right)b}} \right)
\end{equation}
where $n=0,1,2,3,...$ and $N=-1,0,1,2,3,...$. The exact
eigenfunctions can be derived from the following generator:
\begin{equation}\label{ef}
y_n (x) = C_2 \exp \left( { - \int\limits^x {\alpha _k dt} }
\right)
\end{equation}
where $n=0,1,2,...$ and $k\geq n$ is the iteration step number.
Using termination condition of the method given by
Eq.(\ref{itercond}) and $\lambda_0$ and $s_0$ determined by
Eq.(\ref{snln}), the eigenfunctions are obtained as follows;
\begin{small}
\begin{eqnarray*}
y_0 (x) & = & 1 \\
y_1 (x) & = & - C_2 (N + 2)\sigma \left( {1-\frac{b\left( {\rho + 1}\right)}{\sigma }x^{N + 2}} \right) \\
y_2 (x) & = & C_2 (N + 2)^2\sigma \left( {\sigma + 1}
\right)\left( {1 - \frac{2b\left( {\rho + 2} \right)}{\sigma }x^{N
+ 2} + \frac{b^2\left( {\rho + 2} \right)\left( {\rho + 3}
\right)}{\sigma \left( {\sigma + 1} \right)}x^{2(N + 2)}} \right)
\\
 y_3(x) & = & - C_2\frac{\sigma\left({\sigma+1}\right) \left({\sigma+2}
\right)}{\left({N+2}\right)^{-3}} \left (1-\frac{3b\left( {\rho +
3}\right)} \sigma x^{N+2}  \right. \nonumber \\
& + & \left. \frac{3b^2\left({\rho+3}\right)\left(
{\rho+4}\right)}{\sigma\left({\sigma+1} \right)}x^{2(N+2)}
-\frac{b^3\left({\rho+3}\right)\left({\rho+4}\right)\left({\rho+
5} \right)}{\rho \left( {\rho + 1} \right)\left( {\rho + 2}
\right)}x^{3\left( {N + 2} \right)} \right ) \nonumber \\
\ldots \emph{etc}
\end{eqnarray*}
\end{small}
Finally, the following general formula for the exact solutions
$y_n(x)$ is found as;
\begin{equation}\label{efson}
y_n (x) = \left( { - 1} \right)^nC_2 (N + 2)^n\left( \sigma
\right)_n { }_2F_1 ( - n,\rho + n;\sigma ;bx^{N + 2})
\end{equation}

where $(\sigma )_n $=$\frac{\Gamma ( {\sigma + n} )}{\Gamma
(\sigma) }, \quad \sigma$ = $\frac{2m + N + 3}{N + 2}$ \quad
\mbox{and} \quad $\rho$ = $\frac{( {2m + 1} )b + 2a}{( {N + 2}
)b}$.

\section{Energy Eigenvalues Using NU Method}\label{sec3}
In this section, we show how to solve the Schr\"{o}dinger equation
for a particle in the presence of non-central modified Kratzer
potential by using NU method. The standard Kratzer potential is
defined by $V(r)$ =$ -D
\left({\frac{2a}{r}-\frac{a^2}{r^2}}\right)$. Similar to
\cite{cuneyt,fues,fluge}, the modified Kratzer potential is obtained
by adding a D term to the standard Kratzer potential that is
$V(r)~=~D~{\left(\frac{r-a}{r}\right)}^2$. The new exactly solvable
non-central modified Kratzer potential we examine in this paper is
defined as follows:
\begin{equation}
\label{eq21}
 V(r,\theta ) = D\left( {\frac{r - a}{r}} \right)^2 + {\frac{\beta }{r^2\sin^2 \theta }}  + {\frac{\gamma
\cos \theta }{r^2\sin^2 \theta }}
\end{equation}
where $D$ is the dissociation energy and $a$ is the equilibrium
internuclear separation and $\beta$ and $\gamma$ are strictly
positive constants \cite{cuneyt,fues,fluge}. The first term of
this potential is the modified Kratzer potential, the second and
third terms are the angle dependent parts. Thus, the non-central
modified Kratzer potential is defined as a new potential. For the
$CO$ diatomic molecule, a comparison of the modified Kratzer and
the non-central modified Kratzer potentials is shown in
Figure~\ref{fig1} for different $\beta$ and $\gamma$ values with
$\theta=30^{0}$ for the non-central part. The spectroscopic
parameters of the $CO$ diatomic molecule \cite{data1} are given in
Table~\ref{Table1}.

In the spherical coordinates, the Schr\"{o}dinger equation with
the non-central modified Kratzer potential is

\begin{eqnarray}
\label{eq22}
 - \frac{\hbar ^2}{2\mu }\left[ {\frac{1}{r^2}\frac{\partial }{\partial
r}\left( {r^2\frac{\partial }{\partial r}} \right) +
\frac{1}{r^2\sin \theta }\frac{\partial }{\partial \theta }\left(
{\sin \theta \frac{\partial }{\partial \theta }} \right) +
\frac{1}{r^2\sin ^2\theta }\frac{\partial ^2}{\partial \varphi ^2}}
\right]\psi \nonumber \\
 + \left[ {D\left( {\frac{r - a}{r}} \right)^2 + \frac{\beta
}{r^2\sin ^2\theta } + \frac{\gamma \cos \theta }{r^2\sin ^2\theta
}} \right]\psi = E\psi
\end{eqnarray}
If the spherical total wavefunction as $\psi \left( {r,\theta
,\varphi } \right) $= $\frac{{R\left( r \right)}}{r}Y(\theta
,\varphi )=U(r)Y(\theta ,\varphi )$ is inserted into Eq.
(\ref{eq22}), the wave equation for the non-central modified
Kratzer potential is separated into variables and the following
equations are obtained:
\begin{equation}
\label{eq24} \frac{d^2R}{dr^2} + \frac{2\mu }{\hbar ^2}\left[ {E -
D\left( {\frac{r - a}{r}} \right)^2} \right]R -
\frac{\lambda}{r^2} R = 0
\end{equation}

\begin{equation}
\label{eq25}
 \frac{d^2\Theta (\theta )}{d\theta ^2} + \cot \theta
\frac{d\Theta (\theta )}{d\theta } + \left[ {\lambda -
\frac{m^2}{\sin ^2\theta } - \frac{2\mu}{\hbar ^2}\left(
{\frac{\beta + \gamma \cos \theta }{\sin ^2\theta }} \right)}
\right]{\kern 1pt} {\kern 1pt} \Theta (\theta ) = 0
\end{equation}

\begin{equation}
\label{eq26}
 \frac{d^2\Phi (\varphi )}{d\varphi ^2} + m^2\Phi
(\varphi ) = 0
\end{equation}
where $m^2$ and ${\lambda}=l(l+1)$ are the separation constants.
The solution of Eq. (\ref{eq26}) is the well-known azimuthal angle
solution. Eqs. (\ref{eq24}) and (\ref{eq25}) are the radial and
the polar-angle equations. They have been examined separately in
Refs. \cite{cuneyt,yasuk}. In this paper, in order for the
completeness of the paper, instead of just quoting their results
and then discussing the effect of this new potential, we briefly
show how to solve these equations by using the Nikiforov-Uvarov
method \cite{Nikiforov} and then show the physical implications of
this new non-central modified Kratzer potential. The radial part
of the Schr\"{o}dinger equation given by equation (\ref{eq24}) can
be written as
\begin{equation}
\label{eq27} \frac{d^2R}{dr^2} + \frac{2\mu }{\hbar ^2r^2}\left[
{\left( {E - D} \right)r^2 + 2Dar - \left( {Da^2 + \frac{\lambda
\hbar ^2}{2\mu }} \right)} \right]R = 0
\end{equation}
This equation can be further arranged as
\begin{equation}
\label{eq28} \frac{d^2R}{dr^2} + \frac{1}{r^2}\left( {\varepsilon
^2r^2 - \xi r - \kappa } \right)R = 0
\end{equation}
with the following abbreviations
\begin{equation}
\label{eq27a} \varepsilon ^2 = \frac{2\mu \left( {E - D}
\right)}{\hbar ^2}\,\,;\,\, - \xi = \frac{4\mu Da}{\hbar
^2}\,\,;\,\,\kappa = \frac{2\mu \left( {Da^2 + \frac{\lambda \hbar
^2}{2\mu }} \right)}{\hbar ^2}
\end{equation}
It is now suitable for a NU solution. It is necessary to compare Eq.
(\ref{eq28}) with Eq. (\ref{eq11}) to find the solution of this
equation. When these equations are compared, we obtain the following
polynomials:
\begin{equation}
\label{eq29}
 \tilde {\tau }(r) = 0\,\,\,,\,\,\,\sigma (r) =
r\,\,\,,\,\,\,\tilde {\sigma }(r) = \varepsilon ^2r^2 - \xi r -
\kappa
\end{equation}
If these polynomials are inserted into Eq. (\ref{eq17}), we get
$\pi$ function as
\begin{equation}
\label{eq30}
 \pi = \frac{1}{2}\pm \sqrt { - 4\varepsilon ^2r^2 +
4r(k + \xi ) + 4\kappa + 1}
\end{equation}
The expression in the square root must be the square of a polynomial
according to the NU method. Thus, we can determine the constant $k$
by using the condition that the discriminant of the square root is
zero, that is,
\begin{equation}
\label{eq31}
 k = - \xi \pm i\varepsilon \sqrt {1 + 4\kappa }
\end{equation}
In view of that, new possible functions for each $k$ are found as
\begin{equation}
\label{eq32}
 \pi = \left\{ {{\begin{array}{*{20}c}
 {\frac{1}{2}\pm \frac{i}{2}\left[ {2r\varepsilon - i\sqrt {1 + 4\kappa } }
\right]{\begin{array}{*{20}c}
 \hfill & {,{\kern 1pt} {\kern 1pt} {\kern 1pt} for} \hfill & {k = - \xi +
i\varepsilon \sqrt {1 + 4\kappa } } \hfill \\
\end{array} }} \hfill \\
 {\frac{1}{2}\pm \frac{i}{2}\left[ {2r\varepsilon + i\sqrt {1 + 4\kappa } }
\right]{\begin{array}{*{20}c}
 \hfill & {,{\kern 1pt} {\kern 1pt} {\kern 1pt} {\kern 1pt} {\kern 1pt} for}
\hfill & {k = - \xi - i\varepsilon \sqrt {1 + 4\kappa } } \hfill \\
\end{array} }} \hfill \\
\end{array} }} \right.
\end{equation}
To obtain the negative derivative of $\tau = \tilde {\tau } + 2\pi$,
we select
\begin{equation}
\label{eq33}
 \pi (r) = \frac{1}{2} - \frac{i}{2}\left(
{2r\varepsilon + i\sqrt {1 + 4\kappa } } \right)
\end{equation}
and
\begin{equation}
\label{eq331} k = - \xi - i\varepsilon \sqrt {1 + 4\kappa }
\end{equation}
Using $\lambdabar = k + {\pi }'$ together with the values $k$ and
$\pi$, $\tau $ and $\lambdabar $ can be respectively obtained as
\begin{equation}
\label{eq34}
 \tau (r) = 1 + \sqrt {1 + 4\kappa } - 2ir\varepsilon
\end{equation}
\begin{equation}
\label{eq37}
 \lambdabar = - \xi - i\varepsilon \sqrt {1
+ 4\kappa } - i\varepsilon
\end{equation}
Another definition of $\lambdabar_n$ is given at Eq.(\ref{eq19}),
\begin{equation}
\label{eq36} \lambdabar_n  = 2in\varepsilon
\end{equation}
and comparing this with Eq.(\ref{eq37}), the exact energy
eigenvalues of the radial part of the Schr\"{o}dinger equation
with the non-central modified Kratzer potential are derived as
\begin{equation}
\label{eigenk}
 E_{nl} = D - \frac{8\mu D^2 a^2 }{\hbar ^2}\left( {1 + 2n + \sqrt {1 +
4\left( {\frac{2\mu Da^2}{\hbar ^2} + l\left( {l + 1} \right)}
\right)} } \right)^{ - 2}
\end{equation}
The energy eigenvalues calculated by this equation for the $CO$
diatomic molecule are shown in the first column of
Table~\ref{Table2} for different $n$ and $l$ values, which are in
agreement with the results of Ref. \cite{cuneyt}.

It is also possible to determine the radial eigenfunction of this
potential by considering Eq.(\ref{eq12}) and using
Eq.(\ref{eq14}):
\begin{equation}
\label{eq45}
 \phi (r) = r^{  1 / 2(1 + \sqrt {1 + 4\kappa } )}e^{
- i\varepsilon r}
\end{equation}
From Eq. (\ref{eq15}) and Eq. (\ref{eq16}), we get
\begin{equation}
\label{eq41}
 y(r) = \frac{B_{nl}}{\rho (r)}\frac{d^n}{dr^n}\left(
{r ^n}{\rho (r)} \right)
\end{equation}
with $\rho (r) = r^{\sqrt {1 + 4\kappa } }e^{ - 2i\varepsilon r}$.
If we get $2i\varepsilon r = \zeta(r)\,\,;\,\,\sqrt {1 + 4\kappa}
= \nu $, the $ R_{nl}$ radial wave function is
\begin{equation}
\label{eq46}
R_{nl} (\zeta ) = B_{nl} \zeta ^{  1 / 2(1 +\nu )}e^{
- \zeta / 2}L_n^\nu (\zeta )
\end{equation}
if the definition of $U_{nl}(r)= R_{nl}(r)/r$ is used, the total
unnormalized radial wavefunction is
\begin{equation}
U_{nl} (\zeta ) = B_{nl} \zeta ^{ - 1 / 2(1 - \nu
)}e^{ - \zeta / 2}L_n^\nu (\zeta )
\end{equation}
where $B_{nl}$ is the normalization constant and it is evaluated
as follows:
\begin{equation}
\label{eq461}
 B_{nl}=\left(\frac{8 \mu Da}{\hbar^2(2n+\nu+1)}\right)^{3/2} \left[\frac{n!}{(2n+ \zeta +1)(n+
 \zeta)!}\right]^{1/2}
\end{equation}
For the non-central modified Kratzer potential, the angle
dependent Schr\"{o}dinger equation is given in  Eq. (\ref{eq25}).
We may also derive the eigenvalues and eigenfunctions of the polar
angle part of the Schr\"{o}dinger equation \cite{yasuk} similar to
the solutions of the radial part. We can write Eq. (\ref{eq25}) by
introducing a new variable  \cite{yasuk}, $x = \cos \theta$, as
follows:
\begin{equation}
\label{eq47} \frac{d^2\Theta (x)}{dx^2} - \frac{2x}{1 -
x^2}\frac{d\Theta (x)}{dx} + \left( {\frac{\lambda (1 - x^2) - m^2
- \frac{2\mu}{\hbar ^2}\left( {\beta + \gamma x} \right)}{(1 -
x^2)^2}} \right)\Theta (x) = 0
\end{equation}
To apply the Nikiforov-Uvarov method, Eq. (\ref{eq47}) is compared
with Eq. (\ref{eq11}) and the following polynomials are obtained:
\begin{equation}
\label{eq48}
 \tilde {\tau } = - 2x,\quad \sigma = 1 -
x^2,\quad \tilde {\sigma } = - \lambda x^2 - \frac{2\mu
\gamma}{\hbar ^2} x + \left( {\lambda - m^2 - \frac{2\mu
\beta}{\hbar ^2} } \right)
\end{equation}
The function $\pi$ is obtained by putting the
above expression in Eq. (\ref{eq17}),
\begin{equation}
\label{eq49} \pi = \pm \sqrt {x^2(\lambda - k) + \frac{2\mu
\gamma}{\hbar ^2} x - (\lambda - m^2 - \frac{2\mu \beta}{\hbar ^2}
- k)}
\end{equation}
Thus, the polynomial of $\pi$ is found in four possible values:
\begin{equation}
\label{eq50} \pi = \pm \left\{ {{\begin{array}{*{20}c}
{\begin{array}{l}
 x\sqrt {\frac{m^2 + \frac{2\mu
\beta}{\hbar ^2} + u }{2}} + \sqrt {\frac{m^2 + \frac{2\mu
\beta}{\hbar ^2} - u}{2}} {\begin{array}{*{20}c}
 \hfill & {,{\kern 1pt} {\kern 1pt} {\kern 1pt} {\kern 1pt} for} \hfill & {k
= \frac{2\lambda - m^2 - \frac{2\mu \beta}{\hbar ^2} }{2} - \frac{1}{2}u} \hfill \\
\end{array} } \\
 \\
 \end{array}} \hfill \\
 {x\sqrt {\frac{m^2 + \frac{2\mu \beta}{\hbar ^2} - u}{2}} + \sqrt {\frac{m^2 + \frac{2\mu \beta}{\hbar ^2} + u}{2}}
{\begin{array}{*{20}c}
 \hfill & {,{\kern 1pt} {\kern 1pt} {\kern 1pt} {\kern 1pt} {\kern 1pt} for}
\hfill & {k = \frac{2\lambda - m^2 - \frac{2\mu \beta}{\hbar ^2} }{2} + \frac{1}{2}u} \hfill \\
\end{array} }} \hfill \\
\end{array} }} \right.
\end{equation}
where $u = \sqrt {(m^2 + \frac{2\mu \beta}{\hbar ^2} )^2 -
(\frac{2\mu \gamma}{\hbar ^2}) ^2}$. For the polynomial of $\tau =
\tilde {\tau } + 2\pi $ which has a negative derivative,
\begin{equation}
\label{52} \tau = - 2\sqrt {\frac{m^2 + \frac{2\mu \beta}{\hbar
^2} - u}{2}} - 2x\left( {1 + \sqrt {\frac{m^2 + \frac{2\mu
\beta}{\hbar ^2} + u}{2}} } \right)
\end{equation}
Using Eq. (\ref{eq18}) and Eq. (\ref{eq19}), following expressions
for the $\lambdabar$ are obtained respectively
\begin{equation}
\label{eq53}
 \lambdabar = \frac{2\lambda -
(m^2 + \frac{2\mu \beta}{\hbar ^2} )}{2} - \frac{1}{2}u - \sqrt
{\frac{m^2 + \frac{2\mu \beta}{\hbar ^2} + u}{2}}
\end{equation}

\begin{equation}
\label{eq54}
 \lambdabar _n = 2n\left( {1 +
\sqrt {\frac{m^2 + \frac{2\mu \beta}{\hbar ^2} + u}{2}} } \right)
+ n(n - 1)
\end{equation}
If Eq. (\ref{eq53}) and Eq. (\ref{eq54}) are equated and the
definition of $\lambda = l (l + 1)$ is used, the $l$ values are
obtained as
\begin{equation}
\label{eq55}
 l = \sqrt {\frac{m^2 + \frac{2\mu \beta}{\hbar ^2} + \sqrt {(m^2 + \frac{2\mu \beta}{\hbar ^2}
)^2 - (\frac{2\mu \gamma}{\hbar ^2}) ^2} }{2}} + n
\end{equation}
If these $l$ values are inserted into the eigenvalues of the
radial part of the Schr\"{o}dinger equation with the non-central
Kratzer potential given by Eq. (\ref{eigenk}), the energy
eigenvalues are found as follows:

{\tiny \begin{eqnarray} \label{eigennc}
 E_{nm} & = & D - {\frac{8\mu D^2 a^2 } {\hbar ^2}}  \nonumber \\
 &\times & \left[ { 1 + 2n +
\sqrt {1 + 4\left( \frac{2\mu Da^2}{\hbar ^2} + \left( \sqrt
{\frac{m^2 + \frac{2\mu \beta}{\hbar ^2} + \sqrt {(m^2 + \frac{2\mu
\beta}{\hbar ^2} )^2 - (\frac{2\mu \gamma}{\hbar ^2}) ^2} }{2}} + n
\right)\left( \sqrt {\frac{m^2 + \frac{2\mu \beta}{\hbar ^2} + \sqrt
{(m^2 + \frac{2\mu \beta}{\hbar ^2} )^2 - (\frac{2\mu \gamma}{\hbar
^2}) ^2} }{2}} + n + 1 \right) \right) }  }\right]^{-2}
\end{eqnarray}}

We can also obtain the wave function of polar angle part of the
Schr\"{o}dinger equation \cite{yasuk}, using $\sigma $ and $\pi $,
\begin{equation}
\label{eq57}
 \phi(x) = \left( {1 - x} \right)^{B + C / 2}\left( {1 +
x} \right)^{B - C / 2}
\end{equation}
\begin{equation}
\label{eq58}
 \rho(x) = \left( {1 - x^2} \right)^B\left( {\frac{1 +
x}{1 - x}} \right)^{ - C}
\end{equation}
\begin{equation}
\label{eq59}
 y_n(x) = B_n \left( {1 - x} \right)^{ - (B + C)}\left(
{1 + x} \right)^{ - (B - C)}\frac{d^n}{dx^n}\left[ {\left( {1 + x}
\right)^{n + B - C}\left( {1 - x} \right)^{n + B + C}} \right]
\end{equation}
where $B = \sqrt {\frac{m^2 + \frac{2\mu \beta}{\hbar ^2} + u}{2}}
$ and $C = \sqrt {\frac{m^2 + \frac{2\mu \beta}{\hbar ^2} - u}{2}}
$. The polynomial solution of $y_n $ is expressed in terms of
Jacobi polynomials which are one of the ortogonal polynomials,
giving $ \approx P_n^{(B + C,{\kern 1pt} {\kern 1pt} B - C)} (x)$.
The corresponding wave functions are found to be
\begin{equation}
\label{eq60}
 \Theta _n (x) = N_n \left( {1 - x} \right)^{(B + C) /
2}\left( {1 + x} \right)^{(B - C) / 2}P_n^{(B + C,{\kern 1pt}
{\kern 1pt} B - C)} (x)
\end{equation}
where $N_n$ is the normalization constant.

\section{\label{sec5}Energy Eigenvalues Using AIM}

In this section, we show how to solve the Schr\"{o}dinger equation
for a particle in the presence of non-central modified Kratzer
potential by using AIM. In the spherical coordinates, the equation
is given by
\begin{equation}\label{aimsch-1}
\left[ {\frac{1}{r^2}\frac{\partial }{\partial r}\left(
{r^2\frac{\partial }{\partial r}} \right) + \frac{1}{r^2\sin
\theta }\frac{\partial }{\partial \theta }\left( {\sin \theta
\frac{\partial }{\partial \theta }} \right) + \frac{1}{r^2\sin
^2\theta }\frac{\partial ^2}{\partial \varphi ^2} +
\frac{2\mu}{\hbar ^2}\left(E-V(r,\theta ,\varphi )
 \right)} \right]\psi
(r,\theta ,\varphi ) = 0
\end{equation}
If one assigns the corresponding spherical total wave function as
$\psi (r,\theta ,\varphi )$=$\frac{1}{r}R(r)Y(\theta,\varphi )$,
 then by selecting $Y(\theta,\varphi)$=$\frac{1}{sin^{1/2}\theta} H(\theta)\Phi
(\varphi )$, the wave equation (\ref{aimsch-1}) for a general
non-central potential is separated into variables and the
following equations are obtained:
\begin{equation}\label{aimsch-2}
\frac{d^2R(r)}{dr^2} + \left(\frac{2\mu}{\hbar ^2}\left( E-V(r)
\right) - \frac{l(l + 1)}{r^2} \right)R(r) = 0
\end{equation}

\begin{equation}\label{aimsch-3}
\frac{d^2H(\theta)}{d\theta ^2} - \left( {\frac{2\mu}{\hbar
^2}V(\theta ) + \frac{m^2}{\sin ^2\theta } - \frac{1}{2} -
\frac{1}{4}\frac{\cos ^2\theta }{\sin ^2\theta } - l(l + 1)}
\right)H(\theta) = 0
\end{equation}

\begin{equation}\label{aimsch-4}
\frac{d^2\Phi (\varphi )}{d\varphi ^2} = - m^2\Phi (\varphi )
\end{equation}
where $m^2$ and $l(l+1)$ are separation constants. For bound
states, we have the boundary conditions $R$(0)=0 and
$R$($\infty$)=0 in Eq.(\ref{aimsch-2}), $H$(0) and $H$($\pi$) are
infinite in Eq.(\ref{aimsch-3}) and
$\Phi(\varphi)$=$\Phi(\varphi+2\pi)$ in Eq.(\ref{aimsch-4}). If we
specialize to the case where $V(\varphi)$=0, the normalized
solution of the Eq.(\ref{aimsch-4}) that provides the boundary
condition is
\begin{equation}
\Phi_{m}(\varphi)=\frac{1}{\sqrt{2\pi}}e^{im\varphi}  ,
\,\,\,\,\,\,\,\ (m=0,\pm1,\pm2,...)
\end{equation}
In this section we consider solutions of radial and
angle-dependent parts of Schr\"{o}dinger equation for the
non-central modified Kratzer potential within the framework of the
asymptotic iteration method.
The radial Schr\"{o}dinger equation for the non-central modified
Kratzer potential can be written as,
\begin{equation}
\frac{d^2R(r)}{dr^2} + \left(\frac{2\mu}{\hbar ^2}\left( E-D
\left(1-{\frac{2a }{r} + \frac{a ^2}{r^2}} \right) \right) -
\frac{l(l + 1)}{r^2} \right)R(r) = 0 \label{aimsch-r1}
\end{equation}
If the following abbreviations are used:
\begin{equation}
 x = \frac{r}{a }
 \,\,\,\,\,\ \epsilon ^2 = - \frac{2\mu a ^2}{\hbar ^2}(E-D) \,\,\,\,\,\
 \alpha ^2 = \frac{2\mu a ^2}{\hbar ^2}D
\label{aimsch-r2}
\end{equation}
The radial Schr\"{o}dinger equation takes the following form which
is convenient in order to apply AIM.
\begin{equation}
 \frac{d^2R(x)}{dx^2} + \left( { - \epsilon ^2 + \frac{2\alpha ^2}{x} -
\frac{\alpha ^2 + l(l + 1)}{x^2}} \right)R(x) = 0
\label{aimsch-r3}
\end{equation}

\begin{equation}
 \frac{d^2R(x)}{dx^2} + \left( { - \epsilon ^2 + \frac{2\alpha ^2}{x} -
\frac{\tau (\tau - 1)}{x^2}} \right)R(x) = 0 \label{aimsch-r4}
\end{equation}
where
\begin{equation}
 \tau = \frac{1}{2} + \sqrt {\alpha ^2 + \left( {l + \frac{1}{2}}
\right)^2} \,\,\,\,\,\,\
 \tau (\tau - 1) = \alpha ^2 + l(l + 1)
\label{aimsch-r5}
 \end{equation}
In order to solve this equation with AIM for $l\neq0$, we should
transform this equation to the form of Eq.(\ref{diff}). Therefore,
the reasonable physical wave function we propose is as follows
\begin{equation}
 R(x) = x^\tau e^{ - \epsilon x}F(x)
\label{aimsch-r6}
 \end{equation}
If we insert this wave function into the Eq.(\ref{aimsch-r4}), we
have the second-order homogeneous linear differential equations in
the following form
 \begin{equation}
 \frac{d^2F(x)}{dx^2} = 2\left( {\epsilon - \frac{\tau }{x}}
\right)\frac{dF(x)}{dx} + 2\left( {\frac{\epsilon \tau - \alpha
^2}{x}} \right)F(x) \label{aimsch-r7}
\end{equation}
which is now amenable to an AIM solution. To apply the AIM, it is
required to compare Eq.(\ref{aimsch-r7}) with Eq.(\ref{diff1}).
Subsequently, by using Eq. (\ref{iter}), the values of
$\lambda_n(x)$ and $s_n(x)$ are computed as follows
\begin{eqnarray}\label{aimsch-r8}
\lambda _0(x) & = & 2(\epsilon - \frac{\tau }{x }) \nonumber \\
 s_0(x) & = & \frac{2}{x }\left( {\epsilon \tau - \alpha ^2} \right) \nonumber \\
 \lambda _1(x) & = & \frac{2\tau }{x ^2} + \frac{2}{x }\left(
 {\epsilon\tau
- \alpha ^2} \right) + \left( {2\epsilon - \frac{2\tau }{x }} \right)^2 \nonumber \\
s_1(x) & = & 2\left( {\epsilon \tau - \alpha ^2} \right)\left(
{\frac{2\epsilon }{x
} - \frac{2\tau + 1}{x ^2}} \right) \nonumber \\
 \lambda _2(x) & = & \frac{4}{x ^2}\left( {\alpha ^2 - \epsilon \tau -
\frac{\tau }{x }} \right) + \left( {2\epsilon - \frac{2\tau }{x }}
\right)\left[ {\frac{6\tau }{x ^2} + \frac{4(\epsilon \tau -
\alpha ^2)}{x } + \left( {2\epsilon - \frac{2\tau
}{x ^2}} \right)^2} \right] \nonumber \\
 s_2(x) & = & \frac{4\epsilon \tau - 2\alpha ^2}{x ^3} + \frac{2\epsilon \tau -
2\alpha ^2}{x }\left[ {\frac{6\tau }{x ^2} + \frac{2(\epsilon
(\tau - 1) - \alpha ^2)}{x } + \left( {2\epsilon - \frac{2\tau }{x
}} \right)^2}
\right] \\
\ldots \emph{etc} \nonumber
\end{eqnarray}
If we use the termination condition of the AIM given in
Eq.(\ref{itercond}), energy eigenvalues are obtained as follows
\begin{eqnarray}
 \frac{s_0 }{\lambda _0 } = \frac{s_1 }{\lambda _1 }\,\,\,\,\,\,
\Rightarrow
\,\,\,\,\,\,\ \epsilon_0  =  \frac{\alpha ^2}{\tau } \\
 \frac{s_1 }{\lambda _1 } = \frac{s_2 }{\lambda _2 }\,\,\,\,\,\, \Rightarrow
\,\,\,\,\,\,\ \epsilon_1 =  \frac{\alpha ^2}{\tau + 1} \\
 \frac{s_2 }{\lambda _2 } = \frac{s_3 }{\lambda _3 }\,\,\,\,\,\, \Rightarrow
\,\,\,\,\,\,\ \epsilon_2 =  \frac{\alpha ^2}{\tau + 2} \\
\ldots \emph{etc} \nonumber
\end{eqnarray}
which can be generalized as
\begin{equation}\label{aimsch-en1}
\epsilon_n=  \frac{\alpha ^2}{\tau + n} \\
\end{equation}
If one inserts the values of $\epsilon$, $\tau$, and $\alpha$ into
equation (\ref{aimsch-en1}), the ro-vibrational energy spectrum of
the Schr\"{o}dinger equation with the non-central modified Kratzer
potential becomes
\begin{equation}\label{aimsch-en2}
E_{nl} =D - \frac{8\mu a^2 }{\hbar ^2}D ^2\left[ {1+2n + \sqrt {1 +
4\left(\frac{2\mu a^2 D }{\hbar ^2} + l (l+1)\right) }}
\right]^{ - 2} \\
\end{equation}
where $n$ and $l$ are the vibrational and the rotational quantum
numbers, respectively.

As indicated in Section \ref{aim}, we can construct the
corresponding eigenfunctions by using Eq.(\ref{ef}).
Eq.(\ref{aimsch-r7}) provides Eq.(\ref{diff1}) for $N$=$-1$,
$b$=$0$, $a$=$\epsilon$ and $m$=$\tau$-1.

Equation (\ref{efson}) can be written using limit relation as $b
\to 0$ in the following form
\begin{equation}
\lim_{b \to 0} { }_2F_1 ( - n,1/b+a;c ;zb)={ }_1F_1 ( - n;c;z)
\end{equation}
\begin{equation}\label{efson1}
y_n (x) = \left( { - 1} \right)^nC_2 (N + 2)^n\left( \sigma
\right)_n { }_1F_1 ( - n,\sigma ;\frac{2a}{N+2}x^{N + 2})
\end{equation}
Directly solutions $F_n(x)$ for the eigenvalue problem
Eq.(\ref{aimsch-en2}) can be obtained from Eq.(\ref{efson1}) with
the substitution $N$=$-1$ and $a$=$\epsilon$.
\begin{equation}
F_n (x) = \left( { - 1} \right)^nC_2(\sigma)_n~{ }_1F_1 \left( { -
n,\sigma ;2\epsilon x} \right)
\end{equation}
where $\sigma$=$2\tau$, and $\left( \sigma \right)_n =
\frac{\Gamma \left( {\sigma + n}
\right)}{\Gamma \left( \sigma \right)}$. 
Consequently, the radial eigenfunction for Schrodinger equation
with modified Kratzer potential is
\begin{equation}
R(x) = x^{\tau}e^{-\epsilon x}\left( { - 1} \right)^nC_2
(\sigma)_n~{ }_1F_1 \left( { - n,2\tau ;2\epsilon x}\right)
\end{equation}
\begin{equation}
R(r) = (\frac{r}{a})^{\tau}e^{-\epsilon\frac{r}{a}}\left( { - 1}
\right)^nC_2(\sigma)_n~{ }_1F_1 \left( { - n,2\tau
;2\epsilon\frac{r}{a}}\right)
\end{equation}
where $C_2$ is normalization constant.

We now probe solutions of the angle-dependent Schr\"{o}dinger
equation with the non-central modified Kratzer potential using the
same approach. The non-relativistic angular motion of a diatomic
molecule of mass $\mu$ is described by the following equation,
\begin{equation}
\frac{d^2H(\theta)}{d\theta ^2} - \left( {\frac{2\mu}{\hbar
^2}V(\theta ) + \frac{m^2}{\sin ^2\theta } - \frac{1}{2} -
\frac{1}{4}\frac{\cos ^2\theta }{\sin ^2\theta } - l(l + 1)}
\right)H(\theta) = 0 \label{aimsch-a1}
\end{equation}
If we define as $l$=$l^\prime$-$\frac{1}{2}$ and
$l(l+1)$=${l^\prime}^2$-$\frac{1}{4}$ , Eq. (\ref{aimsch-a1}) can
be written as
\begin{equation}
\frac{d^2H(\theta )}{d\theta ^2} - \left[
{\frac{{2\mu}\beta/\hbar^2 + m^2 - 1 / 4}{\sin ^2\theta } +
\frac{{ 2\mu}\gamma cos\theta/\hbar^2}{\sin^2\theta }}
\right]H(\theta ) = -{l^\prime}^2H(\theta) \label{aimsch-a2}
\end{equation}
This equation can be further arranged as
\begin{equation}
\frac{d^2H(\theta )}{d\theta ^2} - \left[
{\frac{\kappa^2+\eta^2-1/4}{\sin ^2\theta } +
\frac{2\kappa\eta\cos\theta}{\sin ^2\theta }} \right]H(\theta ) =
-{l^\prime}^2H(\theta ) \label{aimsch-a3}
\end{equation}
with
\begin{equation}
\kappa^2 = \frac{1}{2}\left[\frac{2\mu \beta }{\hbar ^2} + m^2 +
\sqrt {(m^2 +
\frac{2\mu \beta }{\hbar ^2})^2 -\left(\frac{2\mu \gamma}{\hbar ^2}\right)^2}\right]  \\
\end{equation}

\begin{equation}
 \eta^2 = \frac{1}{2}\left[\frac{2\mu \beta }{\hbar ^2} + m^2 - \sqrt {(m^2 +
\frac{2\mu \beta }{\hbar ^2})^2 - \left(\frac{2\mu \gamma}{\hbar ^2}\right)^2}\right]   \\
\end{equation}
Equation (\ref{aimsch-a3}) can be written in the following form by
introducing a new variable of the form $x=cos\theta$,
\begin{equation}\label{aimsch-a4}
\frac{d^2H(x)}{dx^2} + \frac{x}{1-x^2}\frac{dH(x)}{dx}- \left[
{\frac{\kappa^2+\eta^2+2\kappa\eta x}{(1-x^2)^2}}\right]H(x) =
-\frac{{l^\prime}^2}{1-x^2}H(x)
\end{equation}
Let the angular wave function be factorized as:
\begin{equation}
H(x)=(1-x)^{(\frac{\kappa+\eta}{2}+1/4)}(1+x)^{(\frac{\kappa-\eta}{2}+1/4)}f(x)
\end{equation}
Equation (\ref{aimsch-a4}) reduces to the second-order homogeneous
linear differential equation in the following form
\begin{equation}\label{aimsch-a5}
\frac{d^2f(x)}{dx^2} =2\left(
\frac{\eta+(\kappa+1)x}{1-x^2}\right)\frac{df(x)}{dx}+ \left(
{\frac{(\kappa+1/2)^2-{l^\prime}^2}{1-x^2}}\right)f(x)
\end{equation}
which is convenient to a AIM solution. In order to find the
solution of this equation, it is necessary to compare
Eq.(\ref{aimsch-a5}) with Eq.(\ref{diff}). By means of Eq.
(\ref{iter}), the values of $\lambda_n(x)$ and $s_n(x)$ are
obtained as follows
\begin{eqnarray}\label{aimsch-a6}
\lambda _0(x) & = & \frac{2\eta+\left( {2\kappa + 2} \right)x}{1 -
x^2} \nonumber \\
 s_0(x) & = & \frac{\left( {\kappa + 1/2} \right)^2 - {l^\prime}^2}{1 - x^2}
 \nonumber \\
 \lambda _1(x) & = & \frac{\left( {\kappa + 1 / 2} \right)^2 - {l}'^2 + 2\kappa +
2}{1 - x^2} + \frac{\left( {2\eta + (2\kappa + 2)x} \right)\left(
{2\eta + (2\kappa + 4)x} \right)}{\left( {1 - x^{2}} \right)^2} \nonumber \\
s_1(x) & = & \frac{(\left( {\kappa + 1 / 2} \right)^2 -
{l}'^2)\left( {2\eta + (2\kappa + 4)x} \right)}{\left( {1 - x^2}
\right)^2} \nonumber \\
\lambda _2(x) & = & \frac{2(\eta + (\kappa + 1)x)((\kappa + 1 /
2)^2 - {l}'^2 + 6\kappa + 8)}{(1 - x^2)^2} \nonumber \\ && +
\frac{2(\eta + (\kappa + 1)x)((\kappa + 1 / 2)^2 - {l}'^2 + 8x^2 +
(2\eta + (2\kappa + 2)x)(2\eta + (2\kappa + 8)x))}{(1
- x^2)^3}\nonumber \\
s_2(x) & = & \frac{((\kappa + 1 / 2)^2 - {l}'^2)((\kappa + 1 /
2)^2 - {l}'^2 + 2(2\kappa + 2))}{(1 - x^2)^2} \nonumber \\ && +
\frac{((\kappa + 1 / 2)^2 - {l}'^2)(8x^2 +
6(2\eta + (2\kappa + 2)x)x + (2\eta + (2\kappa + 2)x)^2)}{(1 - x^2)^3}  \\
\ldots \emph{etc} \nonumber
\end{eqnarray}

Combining these results with the condition given by
Eq.(\ref{itercond}) yields
\begin{eqnarray}
 \frac{s_0 }{\lambda _0 } & = & \frac{s_1 }{\lambda _1 }\,\,\,\,\,\, \Rightarrow
\,\,\,\,\,\,\ {l^\prime}^2 = (\kappa+ \frac{1}{2})^2 \\
 \frac{s_1 }{\lambda _1 } & = & \frac{s_2 }{\lambda _2 }\,\,\,\,\,\, \Rightarrow
\,\,\,\,\,\, {l^\prime}^2 = (\kappa + \frac{3}{2})^2 \\
 \frac{s_2 }{\lambda _2 } & = & \frac{s_3 }{\lambda _3 }\,\,\,\,\,\, \Rightarrow
\,\,\,\,\,\, {l^\prime}^2 = (\kappa + \frac{5}{2})^2 \\
\ldots \emph{etc} \nonumber
\end{eqnarray}
When the above expressions are generalized, the eigenvalues turn
out as
\begin{equation}\label{aimsch-a7}
{l^\prime}^2 =  \left( \kappa + N + \frac{1}{2} \right)^2,
\,\,\,\,\,\, N=0,1,2,..
\end{equation}
Inserting $\kappa$ and ${l^\prime}$ in Eq.({\ref{aimsch-a7}})

\begin{equation}\label{aimsch-a8}
l = \sqrt {\frac{\frac{2\mu \beta }{\hbar ^2} + m^2 + \sqrt {(m^2
+\frac{2\mu \beta }{\hbar ^2})^2 - (\frac{2\mu \gamma}{\hbar ^2})^2}
}{2}}+ N
\end{equation}

If $l$ value obtained by Eq.({\ref{aimsch-a8}}) is inserted into
energy spectrum of radial part of the Schr\"{o}dinger equation
given by Eq.(\ref{aimsch-en2}), we find energy spectrum for a
diatomic molecule system in the presence of non-central modified
Kratzer potential as following,

{\tiny \begin{eqnarray} \label{eigenncaim}
 E_{nm} & = & D - {\frac{8\mu D^2 a^2 } {\hbar ^2}}  \nonumber \\
 &\times & \left[ { 1 + 2n +
\sqrt {1 + 4\left( \frac{2\mu Da^2}{\hbar ^2} + \left( \sqrt
{\frac{m^2 + \frac{2\mu \beta}{\hbar ^2} + \sqrt {(m^2 + \frac{2\mu
\beta}{\hbar ^2} )^2 - (\frac{2\mu \gamma}{\hbar ^2}) ^2} }{2}} + N
\right)\left( \sqrt {\frac{m^2 + \frac{2\mu \beta}{\hbar ^2} + \sqrt
{(m^2 + \frac{2\mu \beta}{\hbar ^2} )^2 - (\frac{2\mu \gamma}{\hbar
^2}) ^2} }{2}} + N + 1 \right) \right) }  }\right]^{-2}
\end{eqnarray}}

This eigenvalue equation obtained by using the asymptotic iteration
method is the same as the \ref{eigennc}. As similar to radial
wavefunction, angular wavefunction can be found by using
Eq.(\ref{ef}). The advantage of the asymptotic iteration method is
that it gives the eigenvalues directly by transforming the
second-order differential equation into a form of ${y}''$ =$ \lambda
_0 (r){y}' + s_0 (r)y$. The wave functions are easily constructed by
iterating the values of $s_0(r)$ and $\lambda_0(r)$. The method
presented in this study is a systematic one and puts no constraint
on the potential parameter values involved.

\section{Interpretation of the Results}
In comparison with the eigenvalue equation (\ref{eigenk}) of the
modified Kratzer potential, the eigenvalue equations given by
equations (\ref{eigennc} and \ref{eigenncaim}) have the correction
due to the angle-dependent (non-central) part of the modified
Kratzer potential, given by equation (\ref{eq21}). This correction
has two dependencies: First one is due to the values of the
constants $\beta$ and $\gamma$ in the non-central part of the
potential. Different values of the $\beta$ and $\gamma$ change the
shape and the depth of the potential which affects the values of the
energy eigenvalues. The second one is because of the orbital angular
momentum quantum number $l$ given by equation (\ref{eq55}). It is
known that the orbital angular momentum quantum number $l$ is a good
quantum number and it is a constant of motion for the central
potentials such as the modified Kratzer potential. On the other
hand, for the non-central modified Kratzer potential, it is no
longer a constant of motion and its value is determined in terms of
the constants $\beta$ and $\gamma$ as well as $n$ and $m$ quantum
numbers as given in equation (\ref{eq55}). This is the effect of the
non-central part of the potential which modifies the $l$ values. For
$\beta$=$\gamma$=0, the non-central modified Kratzer potential
reduces to the modified Kratzer and they should give the same energy
eigenvalues for the same $l$ values. For the modified Kratzer
potential, $l$ is a constant of the motion and takes the values
$l$=0,1,2..., but for the non-central modified Kratzer potential, it
is determined from equation (\ref{eq55}). Therefore, they give the
same energy eigenvalues as long as the following condition is
satisfied: the $l$ value of the modified Kratzer potential is equal
to $(n+m)$ of the non-central modified Kratzer potential  for
$\beta$=$\gamma$=0. By comparing the eigenvalue equation ($E_{nm}$)
for the non-central Kratzer potential given by equations
(\ref{eigennc}  and \ref{eigenncaim}) with the eigenvalue equation
($E_{nl}$) for the modified Kratzer potential given by equation
(\ref{eigenk}), we note that the effect of the non-central part is
small for the small values of $\beta$ and $\gamma$ constants, but
the difference becomes apparent when they have large values. In
order to show this, the difference between the eigenvalues of the
modified Kratzer and non-central modified Kratzer potentials is
demonstrated in comparison with each other in Table~\ref{Table2} for
different $n$, $l$ and $m$ quantum numbers.

\section{Conclusion} \label{conc}
In this study, we have calculated the exact bound-state energy
eigenvalues and the corresponding eigenfunctions of the new exactly
solvable non-central modified Kratzer potential by using two
different methods. Both NU and AIM methods generate the same
results. In comparison with the NU method, AIM puts no constraint on
the potential parameter values involved. Therefore, AIm is more
systematic than the NU method in solving such second-order
differential equations.

\begin{figure}
\includegraphics[width=1.0\textwidth]{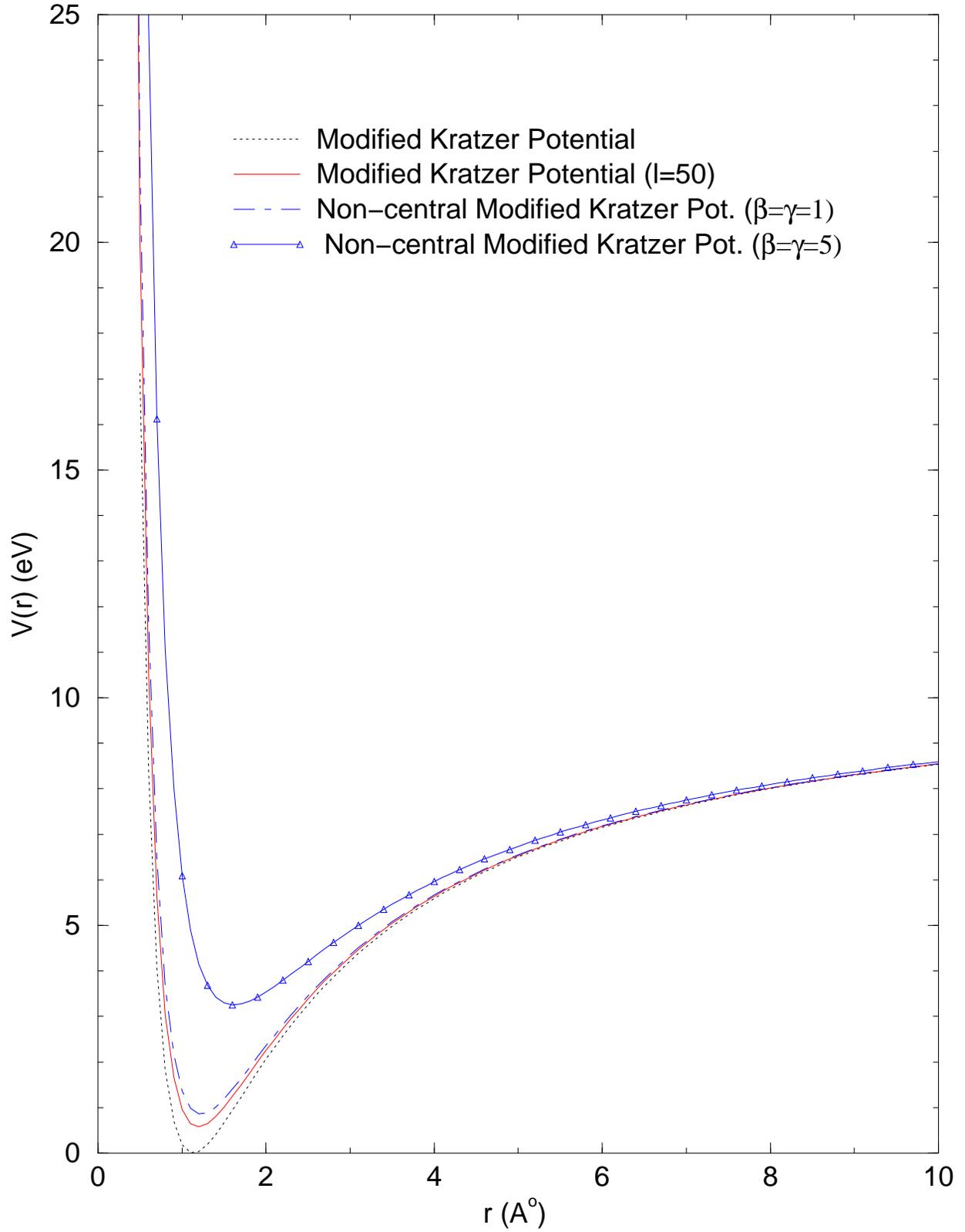} \vskip-0.0cm
\caption{Comparison of the modified Kratzer and non-central modified
Kratzer potentials ($\theta=30^{0}$) for different $\beta$ and
$\gamma$ values for the $CO$ diatomic molecule.} \label{fig1}
\end{figure}

In our study, we have determined the effect of the non-central term
on the bound-state energy eigenvalues explicitly. The energy
eigenvalue equations for the modified and non-central modified
Kratzer potentials are given by equation (\ref{eigenk}) and
equations (\ref{eigennc}  and \ref{eigenncaim}), respectively. In
comparison with the eigenvalues of the modified Kratzer potential,
the effect due to the angle-dependent part of the non-central
potential on the energy eigenvalues is shown for the $CO$ molecule
in Table~\ref{Table2}. The correction is very small for the small
values of $\beta$ and $\gamma$ in the angle dependent part of the
potential. However, the difference becomes apparent for the values
of $\beta$=$\gamma$=1 and over. In order to examine this effect, we
have plotted the shape of the modified Kratzer and non-central
modified Kratzer potential in Figure~\ref{fig1} for different values
of $\beta$ and $\gamma$. We perceive from this figure that the sum
of the modified Kratzer and centrifugal potentials with $l$=50 gives
almost the same shape for the non-central modified Kratzer potential
with $\beta$=$\gamma$=1. That is, the angle-dependent part behaves
like the centrifugal barrier and as it is clearly seen in
Figure~\ref{fig1}, it reduces the depth of the potential pocket,
which effects the bound-state energy eigenvalues.

\section*{Acknowledgments}
This paper is an output of the project supported by the Scientific
and Technical Research Council of Turkey (T\"{U}B\.{I}TAK), under
the project number TBAG-2398 and Erciyes University (FBA-03-27,
FBT-04-15, FBT-04-16).

\begin{table}[h]
\begin{center}
\begin{tabular}{cccc}     \hline     \hline
Molecule  &  D$_{e}$(eV)  &  $a$ (in A$^{0}$) &       $\mu$ (in amu)
\\ \hline
$CO$  & 10.84514471& 1.1282 & 6.860586 \\     \hline     \hline
\end{tabular}
\end{center}
\caption{Reduced mass and spectroscopical properties of the $CO$
diatomic molecule in the ground state \cite{data1}.} \label{Table1}
\end{table}
\begin{table}
\caption{Comparison of the eigenvalues of the modified Kratzer
($E{_{nl}}_{(K)}$) and non-central modified Kratzer
($E{_{nm}}_{(NC-K)}$) potentials for different $n$, $l$ and $m$
values with $\beta=\gamma$=0.0, 0.1, 1.0 and 5.0 values for the $CO$
diatomic molecule, calculated by using equations (\ref{eigenk}),
(\ref{eigennc}) and (\ref{eigenncaim}).}
\begin{center}
\begin{tabular}{ccccccccccc}
\hline\hline
$n$ & & $l$ & & $m$ & & $E{_{nl}}_{(K)}$ & & $E{_{nm}}_{(NC-K)}$ \\
& & & & && & $\beta=\gamma=0.0$ &$\beta=\gamma=0.1$ &$\beta=\gamma=1.0$ &$\beta=\gamma=5.0$ \\
  0 & &    0 & &    0 & &   0.050753 & 0.050753 &   0.092637 &   0.436157 &   1.717889\\
    & &    1 & &    1 & &   0.051227 & 0.051227 &   0.095893 &   0.445310 &   1.733400\\
    & &    2 & &    2 & &   0.052175 & 0.052175 &   0.099399 &   0.454672 &   1.749028\\
    & &    3 & &    3 & &   0.053596 & 0.053596 &   0.103170 &   0.464246 &   1.764773\\
  1 & &    0 & &    0 & &   0.151080 & 0.151547 &   0.198791 &   0.549209 &   1.826511\\
    & &    1 & &    1 & &   0.151547 & 0.152482 &   0.202234 &   0.558423 &   1.841886\\
    & &    2 & &    2 & &   0.152482 & 0.153883 &   0.205932 &   0.567844 &   1.857376\\
    & &    3 & &    3 & &   0.153883 & 0.155751 &   0.209899 &   0.577475 &   1.872981\\
  2 & &    0 & &    0 & &   0.250015 & 0.251397 &   0.303835 &   0.660838 &   1.933480\\
    & &    1 & &    1 & &   0.250475 & 0.252779 &   0.307458 &   0.670109 &   1.948719\\
    & &    2 & &    2 & &   0.251397 & 0.254621 &   0.311342 &   0.679584 &   1.964071\\
    & &    3 & &    3 & &   0.252779 & 0.256923 &   0.315498 &   0.689268 &   1.979537\\
  3 & &    0 & &    0 & &   0.347582 & 0.350309 &   0.407780 &   0.771065 &   2.038826\\
    & &    1 & &    1 & &   0.348037 & 0.352125 &   0.411577 &   0.780387 &   2.053929\\
    & &    2 & &    2 & &   0.348946 & 0.354396 &   0.415639 &   0.789913 &   2.069144\\
    & &    3 & &    3 & &   0.350309 & 0.357118 &   0.419978 &   0.799645 &   2.084470\\
    \hline\hline
\end{tabular}%
\end{center} \label{Table2}
\end{table}
\end{document}